\documentclass[letterpaper,10pt,conference]{ieeeconf}
\IEEEoverridecommandlockouts
\usepackage[T1]{fontenc}
\usepackage{amsmath,amssymb,graphicx,booktabs,cite}
\pdftrailerid{}
\usepackage[hidelinks]{hyperref}
\hypersetup{pdfauthor={},pdftitle={},pdfsubject={},pdfkeywords={},pdfcreator={},pdfproducer={}}
\newtheorem{theorem}{Theorem}
\newtheorem{proposition}{Proposition}
\newtheorem{corollary}{Corollary}
\newtheorem{lemma}{Lemma}

\newcommand{\R}{\mathbb R}
\newcommand{\T}{\mathsf T}
\newcommand{\col}{\operatorname{col}}
\newcommand{\rank}{\operatorname{rank}}
\newcommand{\diag}{\operatorname{diag}}
\newcommand{\Dc}{\mathcal D_c}
\newcommand{\Ds}{\mathcal D}

\title{\LARGE\bf Robust Data-Driven Output Feedback\protect\\for Systems With Unknown Input Delays}
\author{Ngoc Tuan Dinh, Ba Huy Nguyen, Alexey Peregudin, and Egor Dogadin\thanks{Ngoc Tuan Dinh is with ITMO University, St.~Petersburg, Russia (e-mail: \texttt{dinhngoctuan6789@gmail.com}).}\thanks{Ba Huy Nguyen is with Viettel Aerospace Institute, Hoa Lac, Ha Noi, Vietnam (e-mail: \texttt{leningrat206@gmail.com}).}\thanks{Alexey Peregudin is with the School of Electrical and Electronic Engineering, University of Sheffield, Sheffield, United Kingdom (e-mail: \texttt{a.peregudin@sheffield.ac.uk}).}\thanks{Egor Dogadin is with ITMO University, St.~Petersburg, Russia (e-mail: \texttt{egor.dogadin@icloud.com}).}}
\begin{document}
\maketitle
\thispagestyle{empty}
\pagestyle{empty}
\begin{abstract}
In a history-feedback realization, an input delay increases memory without increasing the number of unknown physical coefficients. We use this structure to design output-feedback controllers for discrete-time systems with one input, one output, and known undelayed order, using noisy input--output data. For a known delay, we derive a necessary and sufficient linear matrix inequality for common-quadratic stabilization of all models consistent with the data. Optimizing its equation-error bound gives a robustness margin for the measured record. For an unknown constant delay in a known interval, we characterize noiseless delay ambiguity and give a sufficient condition for one controller to stabilize all models for every delay consistent with the error bound. Under stated excitation and nondegeneracy assumptions, this procedure is feasible for all sufficiently small equation errors. Numerical examples compare robustness at equal controller memory and show that certified stabilization can succeed while distinct delay models remain compatible with the same data.
\end{abstract}

\section{Introduction}
An input delay requires feedback to account for commands still on their way to the actuator. Predictor feedback and Lyapunov--Krasovskii methods address known-model delay problems~\cite{Krstic2009,Fridman2014}. With an unknown plant, an incorrect delay can also be confused with a poor dynamic fit. An unrestricted higher-order fit discards a useful distinction: the delay changes which past inputs enter the output equation, not how many plant coefficients are unknown.

Data-dependent linear matrix inequalities (LMIs) provide stabilizing controllers without explicit model identification~\cite{DePersisTesi2020}. Data informativity studies guarantees over all data-consistent models~\cite{vanWaarde2020}; matrix S-lemma and Petersen-lemma results give exact quadratic tests for specified bounded-error models under suitable regularity assumptions~\cite{vanWaarde2022,Bisoffi2022}. Input--state delay methods use Lyapunov--Krasovskii conditions~\cite{RuedaEscobedo2022}, while~\cite{Portilla2025} combines unknown-delay stabilization with least-squares delay assessment. Output-based unknown-delay regulation in~\cite{Zhao2026} uses continuous-time identification and observer design. Here we certify discrete-time stabilization over all delay descriptions retained by bounded-error data.

Output-feedback designs use measured histories~\cite{Alsalti2025} or autoregressive descriptions~\cite{vanWaarde2024}. Li et al.~\cite{Li2026} already exploit auxiliary-system structure for robust control from noisy input and output data; structural prior information is also treated in~\cite{Berberich2023}. Here each delay candidate has $2n$ uncertain coefficients, with $n$ the undelayed order. We use this input-lag restriction to quantify fixed-record robustness and cover unresolved delays with one controller. Our primary bound concerns equation error; the sensor-noise extension gives a conservative outer enclosure.

The paper makes two contributions.
\begin{enumerate}
\item For a known delay, we derive a necessary and sufficient LMI for common-quadratic stabilization of all data-consistent models (Theorem~\ref{thm:margin}). Optimizing its equation-error budget gives a fixed-record robustness margin. Its centered information radius is homogeneous, monotone, and concave (Corollary~\ref{cor:geometry}). At equal memory and for the same record, the structured fit inherits every certificate for the unrestricted fit, so its margin is no smaller (Proposition~\ref{prop:data}).
\item For an unknown \emph{constant} delay in a known interval, we characterize noiseless ambiguity (Theorem~\ref{thm:identification}) and give a sufficient condition for one controller to stabilize all retained families (Theorem~\ref{thm:unknown}). Under excitation and nondegeneracy assumptions, the complete procedure is feasible for all sufficiently small equation errors (Corollary~\ref{cor:smallnoise}). Unique delay identification~\cite{BjorklundLjung2003} need not precede certified stabilization, as the three-delay example illustrates.
\end{enumerate}
The margin holds the measured record fixed, unlike the experiment-wide tolerance in~\cite{PeregudinDinh2026}. Supporting results explain the obstruction in an unrestricted overparameterized family (Proposition~\ref{prop:overorder}) and enclose sensor errors without a prior coefficient bound (Proposition~\ref{prop:sensor}).

We use $\col$ for column stacking, $*$ for symmetric blocks, $\|\cdot\|_2$ for the Euclidean or spectral norm, and $e_i$ for canonical basis vectors.

\section{Plant, data, and controller memory}
Consider the real, discrete-time plant
\begin{equation}\label{eq:plant}
 x(k+1)=Ax(k)+b u(k-d),\quad y(k)=c^\T x(k),
\end{equation}
where $x\in\R^n$, $u,y\in\R$, and $d\ge0$ is an integer; $u(-d),\ldots,u(-1)$ are initial conditions. The undelayed realization $(A,b,c^\T)$ is minimal and its order $n$ is known. Define
\begin{equation}\label{eq:poly}
\begin{split}
 D(z)&=\det(zI-A)=z^n+\sum_{i=1}^n\alpha_i z^{n-i},\\
 N(z)&=c^\T\operatorname{adj}(zI-A)b=\sum_{i=1}^n\beta_i z^{n-i}.
\end{split}
\end{equation}
Minimality makes $D$ and $N$ coprime. The transfer function is $z^{-d}N(z)/D(z)$, with relative degree $r=\min\{i:\beta_i\ne0\}$. By Cayley--Hamilton, for arbitrary initial states and $k\ge n$,
\begin{equation}\label{eq:arx}
 y(k)=-\sum_{i=1}^n\alpha_i y(k-i)+\sum_{i=1}^n\beta_i u(k-d-i).
\end{equation}
For $k\ge n+d$ only nonnegative input indices occur. Put $\theta=\col(\alpha,\beta)\in\R^{2n}$ and
\begin{equation}\label{eq:regressor}
 \begin{split}
 \varphi_d(k)=\col\big(&-y(k-1),\ldots,-y(k-n),\\
 &u(k-d-1),\ldots,u(k-d-n)\big).
 \end{split}
\end{equation}

\subsection{What the error bound means}
At selected times $\mathcal K=\{k_1,\ldots,k_T\}$, let
\begin{equation}\label{eq:data}
 Y=\theta^\T\Phi_d+W,\quad \|W\|_2^2\le\gamma,
\end{equation}
Here $Y\in\R^{1\times T}$ and $\Phi_d\in\R^{2n\times T}$ collect the columns $y(k_j)$ and $\varphi_d(k_j)$, respectively. Columns may be stacked from several experiments, with valid lag histories in each. The error is an \emph{equation error} in~\eqref{eq:arx}. The bound $\gamma$ must come from a valid disturbance bound; a small least-squares residual alone does not supply one. All synthesis statements concern the disturbance-free plant after data collection.

The compatible coefficient set is
\begin{equation}\label{eq:set}
 \Sigma_d(\gamma)=\{\vartheta:\|Y-\vartheta^\T\Phi_d\|_2^2\le\gamma\}.
\end{equation}
It deliberately allows nonminimal ARX models: retaining them gives a convex uncertainty description that contains the physical plant. If $\Phi_d$ has full row rank, let
\begin{equation}\label{eq:fit}
 H_d=\Phi_d\Phi_d^\T,\mspace{16.100mu}
 \hat\theta_d=H_d^{-1}\Phi_dY^\T,\mspace{16.100mu}
 e_d^2=\|Y-\hat\theta_d^\T\Phi_d\|_2^2.
\end{equation}
Least-squares orthogonality gives
\begin{equation}\label{eq:ellipsoid}
 \Sigma_d(\gamma)=\{\hat\theta_d+\eta:\eta^\T H_d\eta\le\rho_d\},
 \quad \rho_d=\gamma-e_d^2.
\end{equation}
Thus $e_d^2>\gamma$ rejects the candidate, while $\rho_d\ge0$ is the uncertainty budget left after fitting. Formula~\eqref{eq:fit} is analytical; an SVD or QR factorization should be used numerically.

\subsection{A measured history is the controller state}
For $p\ge n$ and $q\ge n+d$, store
\begin{equation}\label{eq:history}
 \begin{split}
 \zeta(k)=\col\big(&y(k-1),\ldots,y(k-p),\\
 &u(k-1),\ldots,u(k-q)\big)\in\R^m,
 \end{split}
\end{equation}
where $m=p+q$. Let $J_\ell$ have ones on its first subdiagonal and zeros elsewhere, and set $A_0=\diag(J_p,J_q)$, $e=e_1\in\R^m$, and $B=e_{p+1}\in\R^m$. Define $S_d\in\R^{2n\times m}$ by $S_d\zeta=\varphi_d$: its first $n$ rows select the first $n$ output entries with a minus sign, and its last $n$ rows select input entries $d+1,\ldots,d+n$. Consequently,
\begin{equation}\label{eq:historydyn}
 \zeta(k+1)=(A_0+e\theta^\T S_d)\zeta(k)+Bu(k).
\end{equation}
The first row computes the next stored output; the other rows only shift measured values. We use $u(k)=K\zeta(k)$, with
\begin{equation}\label{eq:closed}
 A_{d,K}(\vartheta)=A_0+e\vartheta^\T S_d+BK.
\end{equation}
The smallest memory used here is $(p,q)=(n,n+d)$. This is a property of this implementation, not a lower bound on the order of every possible controller.

For example, take $n=2,d=1$ and $(p,q)=(2,3)$. Then $\zeta(k)=\col(y(k-1),y(k-2),u(k-1),u(k-2),u(k-3))$, and the open-loop history matrices are
\[
 A_0+e\theta^\T S_1=
 \begin{bmatrix}
 -\alpha_1&-\alpha_2&0&\beta_1&\beta_2\\
 1&0&0&0&0\\
 0&0&0&0&0\\
 0&0&1&0&0\\
 0&0&0&1&0
 \end{bmatrix},\quad
 B=\begin{bmatrix}0\\0\\1\\0\\0\end{bmatrix}.
\]
Only four entries are unknown. The first-row zero on $u(k-1)$ records the delay; the new command enters row three. A larger delay lengthens the known shifts, leaving four uncertain coefficients.

\begin{lemma}\label{lem:history}
The true history pair $(A_0+e\theta^\T S_d,B)$ is stabilizable. If $A_{d,K}(\theta)$ is Schur, then $x,y,u$ converge exponentially to zero once the history is available and feedback is activated.
\end{lemma}
\begin{proof}
For $p=n,q=n+d$, let $w^\T$ be a left PBH eigenvector orthogonal to $B$, so $w_{u,1}=0$. Its output part is nonzero and obeys the companion recursion, hence $D(\mu)=0$ and $\kappa=w_{y,1}\ne0$. With $g=\kappa[0_d^\T\ \beta^\T]$, the input recursion is $w_{u,j+1}=\mu w_{u,j}-g_j$, with terminal condition $\mu w_{u,q}=g_q$. Therefore $\sum_{j=1}^{q}\mu^{q-j}g_j=\kappa N(\mu)=0$, contradicting coprimeness. Extra memory adds only stable zero modes. Schur stability gives exponential decay of stored outputs and inputs. Observability reconstructs $x(k)$ from the $n$ outputs and required inputs in $\zeta(k+n)$.
\end{proof}

This history also contains the information needed for predictor feedback: reconstruct $x(k-n)$ from the stored outputs, then propagate~\eqref{eq:plant} using the stored inputs to obtain $x(k+d)$. Hence laws $u(k)=F_x x(k+d)$ belong to the class $u=K\zeta$, although the design does not require measured states or an explicit predictor~\cite{Krstic2009}.

\section{Known delay: an exact robustness margin}
Fix the delay and memory. We seek one $K$ and one $Q\succ0$ such that
\begin{equation}\label{eq:quadratic}
 Q-A_{d,K}(\vartheta)Q A_{d,K}(\vartheta)^\T\succ0
 \quad\text{for every }\vartheta\in\Sigma_d(\gamma).
\end{equation}
This is the dual form of a common quadratic Lyapunov certificate; the Lyapunov matrix is $Q^{-1}$. It is stronger than asking each compatible model to be stable separately.

\begin{theorem}\label{thm:margin}
Suppose $\Sigma_d(\gamma)\ne\varnothing$. A certificate~\eqref{eq:quadratic} requires $\rank\Phi_d=2n$. Under this rank condition, it exists if and only if there are $Q=Q^\T\succ0$ and $Z\in\R^{1\times m}$ such that
\begin{equation}\label{eq:raw}
 \begin{split}
 &\mathcal M_d(\gamma,Q,Z):={}\\
 &\begin{bmatrix}
 Q+(YY^\T-\gamma)ee^\T&A_0Q+BZ&eY\Phi_d^\T\\
 *&Q&-QS_d^\T\\
 *&*&H_d
 \end{bmatrix}\succ0.
 \end{split}
\end{equation}
Then $K=ZQ^{-1}$. If the fitted pair $(A_0+e\hat\theta_d^\T S_d,B)$ is stabilizable, define
\begin{equation}\label{eq:gamma}
 \Gamma_d=\sup\{\gamma\ge e_d^2:\exists Q\succ0,Z,\ \mathcal M_d(\gamma,Q,Z)\succ0\}.
\end{equation}
The nonempty sets admitting~\eqref{eq:quadratic} are exactly those with $e_d^2\le\gamma<\Gamma_d$.
\end{theorem}
\begin{proof}
If $\eta^\T\Phi_d=0$ with $\eta\ne0$, any compatible $\vartheta_0$ generates the line $\vartheta_0+t\eta$. The first row of~\eqref{eq:closed} is then $a^\T+t\eta^\T S_d$. Its squared $Q$-norm is unbounded, whereas~\eqref{eq:quadratic} bounds it by $e^\T Qe$. Since $S_d$ has full row rank, this is impossible.

For full rank, write $\hat A_d=A_0+e\hat\theta_d^\T S_d$ and $E_d=\sqrt{\rho_d}H_d^{-1/2}S_d$. By~\eqref{eq:ellipsoid}, the uncertain matrix is $\hat A_d+BK+e\delta^\T E_d$, $\|\delta\|_2\le1$. A Schur complement and the strict Petersen lemma~\cite{Bisoffi2022} give the equivalent condition
\begin{equation}\label{eq:petersen}
 \begin{bmatrix}
 Q-\lambda ee^\T&\hat A_dQ+BZ&0\\
 *&Q&QE_d^\T\\
 *&*&\lambda I_{2n}
 \end{bmatrix}\succ0,\quad\lambda>0.
\end{equation}
For $\rho_d>0$, scale $(Q,Z,\lambda)$ by $\rho_d/\lambda$, then take a Schur complement. This gives
\begin{equation}\label{eq:centered}
 \begin{bmatrix}
 Q-\rho_d ee^\T&\hat A_dQ+BZ&0\\
 *&Q&QS_d^\T\\
 *&*&H_d
 \end{bmatrix}\succ0.
\end{equation}
Eliminating $H_d$ in~\eqref{eq:raw} and~\eqref{eq:centered} gives the same matrix, by~\eqref{eq:fit}. At $\rho_d=0$, nominal stability implies~\eqref{eq:centered} by scaling a nominal Lyapunov matrix sufficiently small; the reverse implication is immediate. Finally, increasing $\gamma$ enlarges the set, and strict feasibility persists under a small increase in $\gamma$. The feasible budgets therefore form the stated interval.
\end{proof}

The significance of~\eqref{eq:gamma} is practical: it measures how much total equation error \emph{this record} can support within the chosen certificate class. The matrices $Y,\Phi_d$ are held fixed while $\gamma$ is optimized. This differs from an experiment-wide guarantee in which the measured record itself changes with the disturbance. Replacing strict inequalities by semidefinite ones gives the same supremum: convex combinations with a strict feasible point approximate any point of the closed feasible set. A boundary optimizer need not be a certificate; recover and verify a controller at a budget below $\Gamma_d$.

\begin{corollary}\label{cor:geometry}
Fix a stabilizable nominal coefficient vector $\theta_0$, delay, and memory. Let $\mathcal R(H)$ be the supremum of radii $\rho$ for which the ellipsoid $\{\theta_0+\eta:\eta^\T H\eta\le\rho\}$ admits~\eqref{eq:quadratic}. For $H\succ0$, $\mathcal R$ is positive, finite, positively homogeneous, monotone in the semidefinite order, and concave. For the fitted nominal model, $\Gamma_d=e_d^2+\mathcal R(H_d)$.
\end{corollary}
\begin{proof}
Use~\eqref{eq:centered} with fixed $\hat A_d$. Convex combinations of feasible tuples prove concavity; scaling all blocks proves homogeneity; increasing $H$ preserves feasibility. Nominal stability gives a positive radius. For finiteness, the ellipsoid contains $\theta_0\pm t e_1$ with $t=\sqrt{\rho/H_{11}}$. Their closed-loop matrices differ by $2t ee^\T$. If both are strict contractions in the same quadratic norm, their half-difference is also a strict contraction, so $t<1$. Thus $\mathcal R(H)\le H_{11}$.
\end{proof}

At a fixed nominal model, more information cannot reduce the certified radius. The concavity statement further quantifies how this radius behaves when information matrices are combined; it does not compare records with different fitted nominal models.

\begin{proposition}\label{prop:data}
For a noiseless experiment, use columns $k=n+d,\ldots,n+d+T-1$. If $u(0),\ldots,u(T+n-2)$ is persistently exciting (PE) of order $2n$, then $\rank\Phi_d=2n$. Here PE of order $\ell$ means full row rank of the depth-$\ell$ input Hankel matrix. Thus $T\ge2n$ is necessary for~\eqref{eq:quadratic}, while $T\ge3n$ permits this sufficient PE condition, independently of $d$.

An unstructured order-$(n+d)$ ARX model instead needs rank $2(n+d)$. At \emph{equal memory} $p=q=n+d$, every common-quadratic certificate for its compatible set is also a certificate for~\eqref{eq:set}.
\end{proposition}
\begin{proof}
Put $v(t)=u(t-d)$. The regressor is an invertible linear image of $\col(x(k-n),v(k-n),\ldots,v(k-1))$: its diagonal blocks are a reversed observability matrix and an input reversal. The state/input rank lemma~\cite[Cor.~2]{Willems2005}, at depth $n$, gives rank $2n$ from PE of order $2n$. The stated input sequence has length $T+n-1$, so its depth-$2n$ Hankel matrix can have rank $2n$ only if $T\ge3n$.

The unstructured regressor contains $n+d$ output and input lags. Its full-rank necessity follows as in Theorem~\ref{thm:margin}. The embedding $\theta\mapsto\col(\alpha,0_d,0_d,\beta)$ preserves residuals and closed-loop matrices at equal memory. Its compatible set therefore contains the structured set.
\end{proof}

\begin{table}[t]
\centering
\caption{Known-delay dimensions. Column bounds are necessary; they do not alone guarantee rank or stabilizability.}
\label{tab:dimensions}
\begin{tabular}{@{}lcc@{}}
\toprule
Quantity & Structured & Unstructured\\
\midrule
Unknown coefficients & $2n$ & $2(n+d)$\\
Necessary columns & $T\ge2n$ & $T\ge2(n+d)$\\
History $(p,q)$ & $(n,n+d)$ & $(n+d,n+d)$\\
LMI order & $6n+2d$ & $6(n+d)$\\
\bottomrule
\end{tabular}
\end{table}

Table~\ref{tab:dimensions} separates parameter count from memory. At equal memory $p=q=n+d$, the structured LMI order is $6n+4d$; its uncertainty still has dimension $2n$. Factoring the data costs $O(n^2T)$ versus $O((n+d)^2T)$, whereas SDP cost also grows with the LMI order. Neither delay-independent parameter count nor PE order means delay-independent duration: the final output is at $n+d+T-1$, spanning $n+d+T$ sampling instants from time zero.

\section{Unknown delay: retain, then stabilize}
Let $\Ds=\{d_{\min},\ldots,d_{\max}\}$ contain the true constant delay, and let $\Delta=d_{\max}-d_{\min}$. Every candidate uses the same columns, with $k_j\ge n+d_{\max}$, and the common memory $p=n,q=n+d_{\max}$. Retain
\begin{equation}\label{eq:retain}
 \Dc(\gamma)=\{j\in\Ds:\min_\vartheta\|Y-\vartheta^\T\Phi_j\|_2^2\le\gamma\}.
\end{equation}
The true delay belongs to this set under~\eqref{eq:data}. The test rejects an explanation only when its best fit already violates the error bound.

\subsection{What noiseless data can distinguish}
For this subsection and the small-error result below, assume additionally that $A$ is nonsingular and $N(0)=\beta_n\ne0$ for the \emph{true physical plant}. The candidate sets~\eqref{eq:set} remain relaxed ARX sets; these extra restrictions are not imposed on every candidate.

\begin{theorem}\label{thm:identification}
Use one noiseless experiment with columns $k=n+d_{\max},\ldots,n+d_{\max}+T-1$. If $u(0),\ldots,u(T+n+\Delta-2)$ is PE of order $2n+\Delta$, then all $\Phi_j$ have rank $2n$, and
\begin{equation}\label{eq:ambiguity}
 \Dc(0)=\{d,\ldots,d+r-1\}\cap\Ds,\quad d=\min\Dc(0).
\end{equation}
The fit at every retained delay $j=d+s$ is unique: $\theta_s=\col(\alpha,\beta_{s+1},\ldots,\beta_n,0_s)$. It satisfies $\theta_s^\T S_{d+s}=\theta^\T S_d$, so these candidates describe the same input--output behavior. The PE condition requires $T\ge3n+\Delta$.
\end{theorem}
\begin{proof}
Set $v(t)=u(t-d)$ and $\Delta_+=d_{\max}-d$, $\Delta_-=d-d_{\min}$. Each output and candidate regressor is a linear function of
\[
 \begin{split}
 \xi(k)=\col\big(&x(k-n-\Delta_+),v(k-n-\Delta_+),\\
 &\ldots,v(k-1+\Delta_-)\big).
 \end{split}
\]
The state/input rank lemma~\cite{Willems2005}, now at depth $n+\Delta$, makes the data matrix of $\xi$ full row rank $2n+\Delta$. Thus any linear identity on the data holds for arbitrary $\xi$.

Set its input entries to zero. Since $A$ is invertible, $x(k-n)$ is arbitrary; observability then makes the output part of any relation $\eta^\T\varphi_j=0$ vanish. The remaining input entries are distinct independent coordinates, so $\eta=0$. This proves the rank claim. Likewise, an exact candidate fit first forces $\alpha'=\alpha$. After canceling these output terms, arbitrary inputs give
\[
 \sum_{i=1}^n\beta_i u(k-d-i)=\sum_{i=1}^n\beta'_i u(k-j-i).
\]
The nonzero extreme terms at lags $d+r$ and $d+n$ require $j+1\le d+r$ and $j+n\ge d+n$, hence $d\le j\le d+r-1$. Conversely, shifting the leading zero numerator coefficients gives exactly the stated $\theta_s$, identically for every history. The PE length count gives $T\ge3n+\Delta$.
\end{proof}

For example, $n=2,r=2$ allows descriptions with delays $d$ and $d+1$: the sole nonzero input term occurs at lag $d+2$ in both. For $s>0$, the shifted numerator is $z^sN(z)$ and has a zero at the origin. Such a description is allowed by the relaxed test, even though the true physical plant has no such zero. The smallest retained delay identifies the physical delay under the stated priors. Without $N(0)\ne0$, an earlier delay can also fit.

\begin{proposition}\label{prop:overorder}
On noiseless data, let $s=d_{\max}-d>0$ and fit an unstructured ARX model of order $h=n+d_{\max}$. Then $\rank\Phi_{\rm ag}\le2h-s$. Moreover, its compatible unreduced history models include a model with an unstable uncontrollable mode. No history feedback stabilizes this entire set.
\end{proposition}
\begin{proof}
For every monic polynomial $p_s$ of degree $s$, the pair $(p_s(z)z^dD(z),p_s(z)N(z))$ gives an exact order-$h$ ARX equation. Its coefficients form an injective affine family of dimension $s$, proving the rank bound. Choose $p_s$ with a real root $\mu>1$. Both fitted polynomials vanish at $\mu$; the PBH recursion in Lemma~\ref{lem:history}, read in reverse, gives an uncontrollable eigenvalue $\mu$ of the unreduced history pair. Feedback cannot move it.
\end{proof}

This obstruction concerns the entire unrestricted fitted family, not the reduced physical plant. Lee et al.~\cite{Lee2026} select an ARX representative with stable nonminimal modes; such a selection does not certify every member of the unrestricted family.

\subsection{One gain for all retained candidates}
Assume each retained $\Phi_j$ has rank $2n$. Form $\hat A_j=A_0+e\hat\theta_j^\T S_j$ and $E_j=\sqrt{\gamma-e_j^2}H_j^{-1/2}S_j$ as before.

\begin{theorem}\label{thm:unknown}
Suppose there are shared matrices $F\in\R^{m\times m}$, $L\in\R^{1\times m}$ and, for every $j\in\Dc(\gamma)$, matrices $P_j=P_j^\T\succ0$ and scalars $\lambda_j>0$ satisfying
\begin{equation}\label{eq:unknown}
 \begin{bmatrix}
 P_j-\lambda_j ee^\T&\hat A_j F+BL&0\\
 *&F+F^\T-P_j&F^\T E_j^\T\\
 *&*&\lambda_j I_{2n}
 \end{bmatrix}\succ0.
\end{equation}
Then $F$ is nonsingular and $K=LF^{-1}$ stabilizes every $\vartheta\in\Sigma_j(\gamma)$ for every retained $j$. Each $P_j^{-1}$ is a common quadratic Lyapunov matrix within its coefficient set. With one candidate, feasibility is equivalent to Theorem~\ref{thm:margin}.
\end{theorem}
\begin{proof}
The middle block implies $F+F^\T\succ P_j\succ0$, so $F$ is nonsingular. Petersen's lemma transforms~\eqref{eq:unknown} into $[P_j\ \ A_{j,K}(\vartheta)F;\ *\ \ F+F^\T-P_j]\succ0$ for every compatible $\vartheta$. Since
\[
 F^\T P_j^{-1}F-(F+F^\T-P_j)
 =(F-P_j)^\T P_j^{-1}(F-P_j)\succeq0,
\]
replacing the lower block by $F^\T P_j^{-1}F$ preserves positivity. A Schur complement gives $P_j-A_{j,K}(\vartheta)P_jA_{j,K}(\vartheta)^\T\succ0$. For one candidate, choose $F=P_j=Q$ and $L=Z$ in~\eqref{eq:petersen} for the converse.
\end{proof}

The shared variables enforce one implementable gain; the $P_j$ may differ. Thus~\eqref{eq:unknown} does not certify arbitrarily switching delays. It also does not claim that infeasibility rules out every simultaneous stabilizing controller. The dilation follows the robust stability principle in~\cite{deOliveira1999}.

The design procedure is therefore direct. First, fix the delay interval, common data columns, and memory. Second, fit each candidate and reject only those whose residual exceeds the prescribed bound. Third, check full row rank for every survivor and solve~\eqref{eq:unknown} jointly; with a single survivor, the exact test~\eqref{eq:raw} is available. Finally, recover $K$, verify strict positivity of the numerical certificate, and activate feedback from the stored history. A consistent rank-deficient candidate prevents a certificate of the present type by Theorem~\ref{thm:margin}; it cannot simply be discarded.

\begin{corollary}\label{cor:smallnoise}
Fix the input, initial histories, and finite experiment in Theorem~\ref{thm:identification}, and let $\Dc^0$ denote the candidate set of its noiseless reference record. During collection, let the ARX recursion have equation error with total energy at most $\gamma$ over \emph{all} generated samples. There exists $\bar\gamma>0$ such that, for every $0\le\gamma\le\bar\gamma$ and every such error, all candidate regressors have full row rank, $\Dc(\gamma)=\Dc^0$, and~\eqref{eq:unknown} is feasible. For a known delay, the same conclusion follows from the rank condition in Proposition~\ref{prop:data} without the extra nonsingularity assumptions.
\end{corollary}
\begin{proof}
Over a fixed finite horizon, the perturbed record differs from the noiseless one by $O(\sqrt\gamma)$, uniformly over these errors. Full row rank therefore persists. Every false noiseless candidate has a strictly positive least-squares residual; continuity and finiteness of $\Ds$ keep all such squared residuals above $\gamma$ for small $\gamma$. Every equivalent candidate remains consistent because $\theta_s^\T S_{d+s}=\theta^\T S_d$ on arbitrary histories: its equation error is the same $W$. For each survivor, $\hat\theta_j=\theta_s+O(\sqrt\gamma)$ and $E_j=O(\sqrt\gamma)$ uniformly. Their nominal history matrices converge to one stabilizable matrix. Choose a strict nominal certificate $Q,K$ using Lemma~\ref{lem:history}, and set $F=P_j=Q$, $L=KQ$, $\lambda_j=\sqrt\gamma$. After eliminating the last block of~\eqref{eq:unknown}, its perturbation tends to zero, so strict positivity persists. At $\gamma=0$, use sufficiently small positive $\lambda_j$. The singleton argument needs only nominal stabilizability and full rank.
\end{proof}

\subsection{Using a sensor-noise bound}
Sensor noise perturbs both $Y$ and the output rows of $\Phi_j$. Its induced equation error is $v(k)+\sum_i\alpha_i v(k-i)$ and therefore depends on unknown coefficients. For sensor error alone, a prior $\|\alpha\|_1\le a_*$ and $|v(k)|\le\bar v$ gives $\gamma=T(1+a_*)^2\bar v^2$. The following alternative needs no coefficient prior; it gives a conservative outer set for this errors-in-variables problem~\cite{Li2026}.

Fix a candidate and suppress its subscript. Write the measured regressor as $\Phi=\col(-X,U)$, where $X$ contains the $n$ output lags and $U$ the candidate input lags. Inputs are exact. Collect every distinct sensor-error sample used by the record into one vector $v$, with $\|v\|_2\le\delta_v$, and let $L_i v$ select the errors at lag $i$, $i=0,\ldots,n$. The same sample has the same coordinate wherever it reappears. Allow additional selected equation errors with norm at most $\delta_w$. Let $U^\dagger$ denote the Moore--Penrose inverse and define
\begin{equation}\label{eq:sensorconstants}
 \begin{split}
 s_\perp&=\sigma_{\min}\big(X(I-U^\dagger U)\big),\mspace{5.700mu}
 \kappa=\|\col(L_1,\ldots,L_n)\|_2,\\
 b_v&=\delta_w+\delta_v\Big\|L_0+\sum_i\hat\alpha_i L_i\Big\|_2,
 \quad \tau=\kappa\delta_v/s_\perp.
 \end{split}
\end{equation}
The projection $I-U^\dagger U$ removes the part of the output histories explained by the input histories; $s_\perp$ measures the independent information that remains.

\begin{proposition}\label{prop:sensor}
Suppose $\Phi$ has full row rank, so $s_\perp>0$, and $\tau<1$. Write $e_{\rm LS}=\|Y-\hat\theta^\T\Phi\|_2$. If $e_{\rm LS}>b_v/\sqrt{1-\tau^2}$, the candidate is inconsistent with the sensor and equation-error bounds. Otherwise, every compatible coefficient vector belongs to
\begin{equation}\label{eq:sensorbound}
 \begin{split}
 &(\vartheta-\hat\theta)^\T\Phi\Phi^\T(\vartheta-\hat\theta)\le\eta_v^2,\\
 &\eta_v=\frac{b_v\tau+\sqrt{b_v^2-(1-\tau^2)e_{\rm LS}^2}}{1-\tau^2}.
 \end{split}
\end{equation}
Equivalently, this is~\eqref{eq:set} for the measured record with $\gamma_{\rm eff}=e_{\rm LS}^2+\eta_v^2$.
\end{proposition}
\begin{proof}
Put $z=\|\Phi^\T(\vartheta-\hat\theta)\|_2$. Projecting onto $\ker U$ gives $\|\alpha-\hat\alpha\|_2\le z/s_\perp$. The measured equation identity is $Y^\T-\Phi^\T\vartheta=w+(L_0+\sum_i\alpha_iL_i)v$, whose norm is at most $b_v+\tau z$. Least-squares orthogonality then gives $e_{\rm LS}^2+z^2\le(b_v+\tau z)^2$. Solving this scalar quadratic proves both claims.
\end{proof}

The condition $\tau<1$ checks whether data separate denominator uncertainty from sensor error sufficiently well. Its failure is inconclusive, not grounds for rejecting a delay. Passing the residual test also does not establish physical consistency:~\eqref{eq:sensorbound} is an outer enclosure. Apply Theorem~\ref{thm:margin} with $\gamma_{\rm eff}$. If full rank and $\tau_j<1$ hold for \emph{every} unrejected candidate, use~\eqref{eq:unknown} with $E_j=\eta_{v,j}H_j^{-1/2}S_j$. The enclosure may be conservative even though subsequent ellipsoid certification is exact at a fixed delay.

\section{Numerical examples}\label{sec:numerics}
Both examples use $D(z)=(z-1.1)(z-0.8)$ and $N(z)=0.2z+0.1$: the undelayed plant is minimal and unstable. Least squares and uncertainty factors use an SVD. We solve the LMIs with CVXPY~1.7.5/Clarabel~0.11.1 and independently reconstruct each returned matrix. A certificate is accepted only when its minimum eigenvalue exceeds $100N\epsilon_{\rm mach}\max(1,\|M\|_2)$ for an $N\times N$ matrix $M$. This floating-point verification uses no interval arithmetic; solver statuses alone are not taken as certificates.

\subsection{Robustness at equal controller memory}
For each $d\in\{1,3,5\}$, twelve length-40 experiments use independent standard Gaussian inputs and equation errors uniform in $[-10^{-3},10^{-3}]$. The same columns $k=8,\ldots,39$ give $T=384$ for both fits. Both controllers have $p=q=2+d$; the structured fit estimates four coefficients and the unstructured fit estimates $2(2+d)$. Initial output pairs are standard Gaussian, corresponding to physical initial states by observability. The seed is 439.

Table~\ref{tab:margin} reports numerical values of~\eqref{eq:gamma}. These are squared equation-error energies for fixed records, not noise-amplitude tolerances across regenerated experiments. Separate strict certificates are recovered at $e_d^2+0.99(\Gamma_d-e_d^2)$; their smallest relative eigenvalue exceeds $2.5\cdot10^{-8}$.

For an independent upper bound, construct an unstructured coefficient vector whose numerator and denominator share $z=-1$. The PBH argument of Proposition~\ref{prop:overorder} makes its full history realization unstabilizable. Its residual energy bounds the unstructured margin from above. The root identities, PBH witness, and residual energies are checked in exact rational arithmetic on the saved record. Table~\ref{tab:margin} rounds these upper bounds $U_{\rm ag}$ upward; dividing the structured strict lower budget by $U_{\rm ag}$ gives the certified ratio. The separation therefore does not depend on boundary optimality or solver failure.

\begin{table}[ht]
\centering
\caption{Matched-memory margins: numerical SDP values, independent upper bounds $U_{\rm ag}$, and certified ratio lower bounds.}
\label{tab:margin}
\begin{tabular}{@{}ccccr@{}}
\toprule
$d$ & $\Gamma_{\rm str}$ & $\Gamma_{\rm ag}$ & $U_{\rm ag}$ & \shortstack{Ratio\\lower bound}\\
\midrule
1 & $4.377$ & $1.879$ & $3.547$ & $1.22$\\
3 & $2.277$ & $6.562\cdot10^{-2}$ & $8.131\cdot10^{-2}$ & $27.7$\\
5 & $1.319$ & $3.080\cdot10^{-3}$ & $3.735\cdot10^{-3}$ & $349.5$\\
\bottomrule
\end{tabular}
\end{table}

\subsection{Three genuinely different delays remain plausible}
Now $d=2$, $\Ds=\{1,2,3\}$, and $(p,q)=(2,5)$. Twelve length-16 experiments use period-two inputs, each alternating between two independent standard Gaussian amplitudes; initial output pairs are also standard Gaussian (seed 806). Selected columns $k=5,\ldots,15$ give $T=132$. Errors uniform in $[-0.01,0.01]$ give the predetermined bound $\gamma=0.0132$.

All three candidates have rank four, condition number $\kappa_2(\Phi_j)=89.44$, residual energy $0.004189$, and centered squared radius $0.009011$. Period-two probing deliberately fails the PE assumption of Theorem~\ref{thm:identification}, so unique delay identification is not expected. The three chosen compatible plants are genuinely different: their first nonzero impulse responses occur at samples 2, 3, and 4.

\begin{figure}[t]
\centering
\includegraphics[width=\columnwidth]{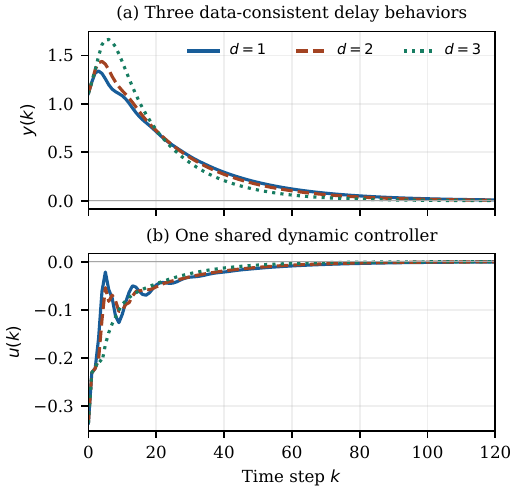}
\caption{One controller stabilizes three genuinely different compatible delayed plants. All use $\zeta(0)=(1,0.9,0,0,0,0,0)^\T$ and activate at $k=0$, with identical controller gains.}
\label{fig:unknown}
\end{figure}

We solve~\eqref{eq:unknown} with $\sum_j\operatorname{tr}P_j=3$. A secondary quadratic gain surrogate retains half the maximum strict margin at this fixed scale. The final minimum ratio $\lambda_{\min}(M_j)/\|M_j\|_2$, for the LMI matrices $M_j$, is $1.05\cdot10^{-4}$. This certifies the entire three-ellipsoid family. Figure~\ref{fig:unknown} illustrates the same gain on three exact compatible plants: $\alpha=(-1.9,0.88)$, $\beta=(0.2,0.1)$ at $d=2$, and $\beta=(0.1,0.2)$ at $d=1,3$. With zero past inputs, observability supplies a physical state for each plant having $y(-2)=0.9,y(-1)=1$. All activate at $k=0$. These simulations illustrate the certificate.

\newpage
\section{Conclusion}
Known delay registers separate feedback memory from a fixed $2n$-dimensional uncertainty. This yields an exact common-quadratic fixed-record margin and a certified shared controller over retained constant-delay families. Uniform small-error feasibility holds under explicit excitation assumptions. Matched-memory bounds demonstrate the structural gain; the three-delay experiment shows that delay identification need not precede certified stabilization. Simultaneous synthesis and sensor-noise enclosure remain conservative.

\end{document}